\documentclass{iau}
\usepackage{graphicx}

\title[Combining magnetic and seismic studies] 
{Combining magnetic and seismic studies to constrain processes in massive stars}

\author[Neiner et al.]   
{Coralie Neiner$^1$,
Pieter Degroote$^{2,1}$, 
Blanche Coste$^{1}$, 
Maryline Briquet$^{3}$
\and St\'ephane Mathis$^{4,1}$
}

\affiliation{$^1$LESIA, UMR 8109 du CNRS, Observatoire de Paris, UPMC, Univ. Paris Diderot, 5 place Jules Janssen, 92195 Meudon Cedex, France\\ email: {\tt coralie.neiner@obspm.fr}\\[\affilskip]
$^2$Instituut voor Sterrenkunde, Celestijnenlaan 200D, B-3001 Heverlee, Belgium\\[\affilskip]
$^3$Institut d'Astrophysique et de G\'eophysique, Universit\'e de Li\`ege, All\'ee du 6 Ao\^ut 17, B\^at B5c, 4000 Li\`ege, Belgium\\[\affilskip]
$^4$Laboratoire AIM Paris-Saclay, CEA/DSM-CNRS-Universit\'e Paris Diderot; IRFU
/SAp, Centre de Saclay, 91191 Gif-sur-Yvette Cedex, France}

\pubyear{2013}
\volume{302}  
\pagerange{119--126}
\jname{Magnetic fields throughout stellar evolution}
\editors{A.C. Editor, B.D. Editor \& C.E. Editor, eds.}
\begin{document}

\maketitle

\begin{abstract}
The presence of pulsations influences the local parameters at the surface of
massive stars and thus it modifies the Zeeman magnetic signatures. Therefore it
makes the characterisation of a magnetic field in pulsating stars more difficult
and the characterisation of pulsations is thus required for the study of
magnetic massive stars. Conversely, the presence of a magnetic field can inhibit
differential rotation and mixing in massive stars and thus provides important
constraints for seismic modelling based on pulsation studies. As a consequence,
it is necessary to combine spectropolarimetric and seismic studies for all
massive classical pulsators. Below we show examples of such combined studies and
the interplay between physical processes.
\end{abstract}

\firstsection 
\section{Modelling oblique magnetic dipoles with pulsations}

$\beta$\,Cep is a magnetic pulsating star. It hosts a radial pulsation mode as
well as non-radial pulsations of lower amplitude. The line profiles in its
spectrum therefore show mainly variations with the pulsation periods in addition
to the Zeeman broadening due to the magnetic field. Measurements of the magnetic
field in Stokes V clearly show both variations due to the pulsations and the
Zeeman signatures of its field. Therefore it is mandatory to take pulsations
into account when trying to determine the magnetic field configuration and
strength from the Stokes profiles.

\begin{figure}[t]
\begin{center}
\includegraphics[height=3cm]{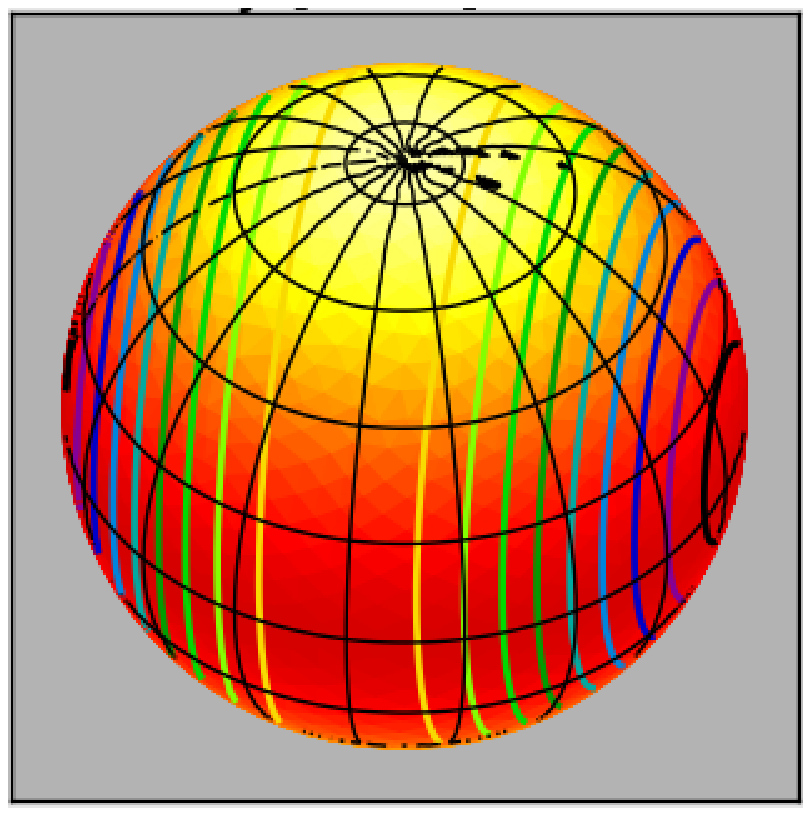} 
\includegraphics[height=3cm]{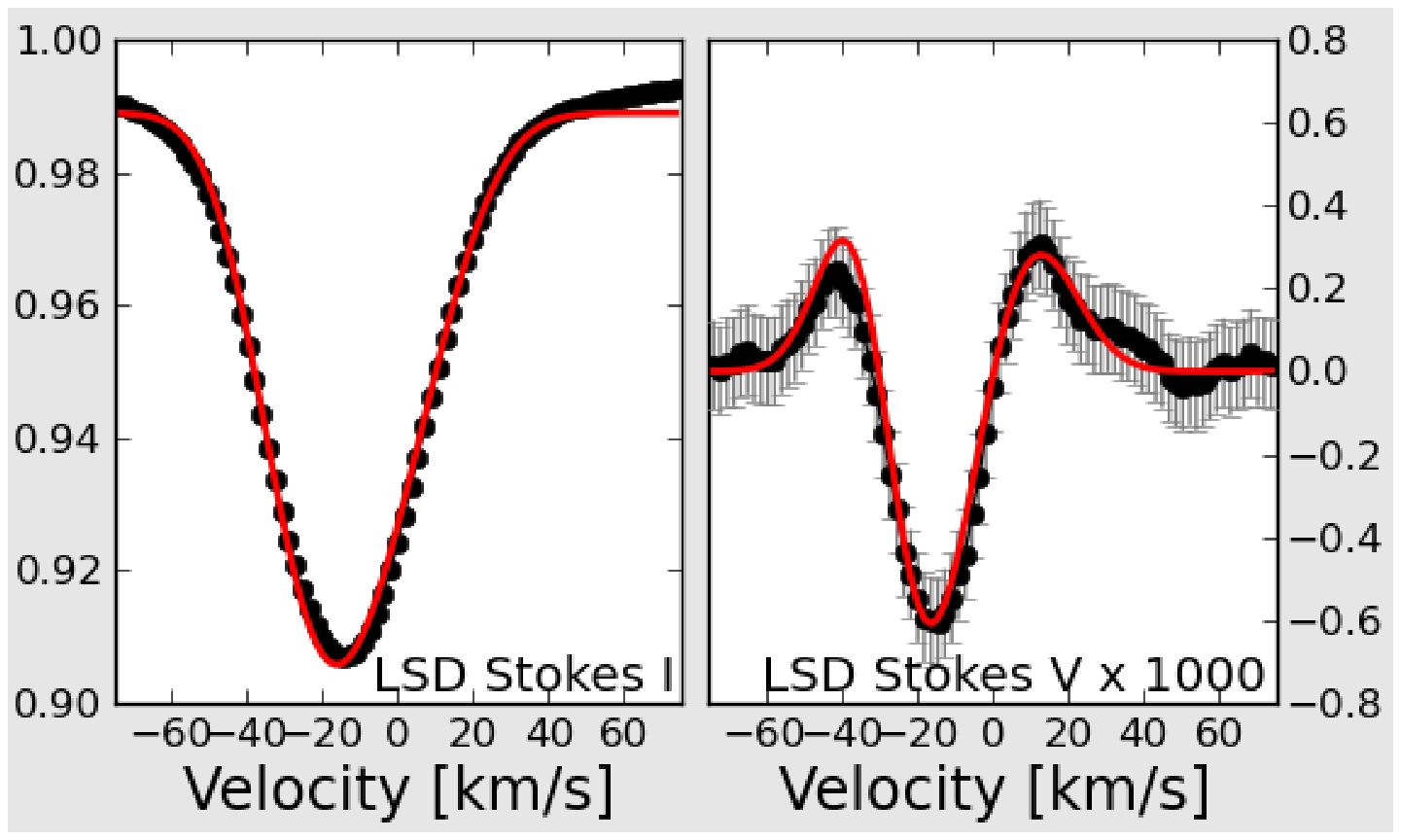}\\
\includegraphics[height=3cm]{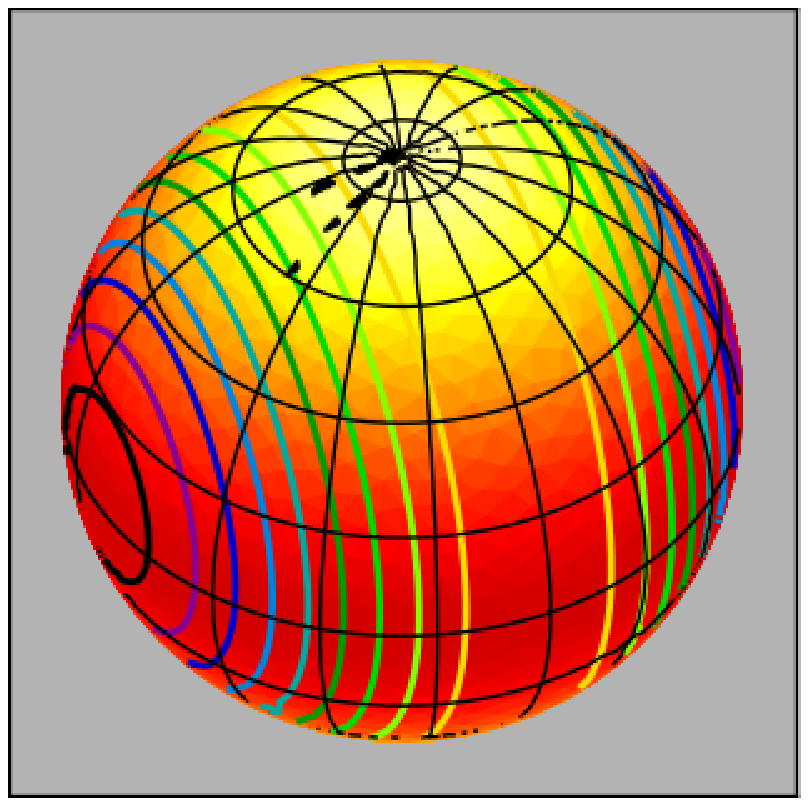} 
\includegraphics[height=3cm]{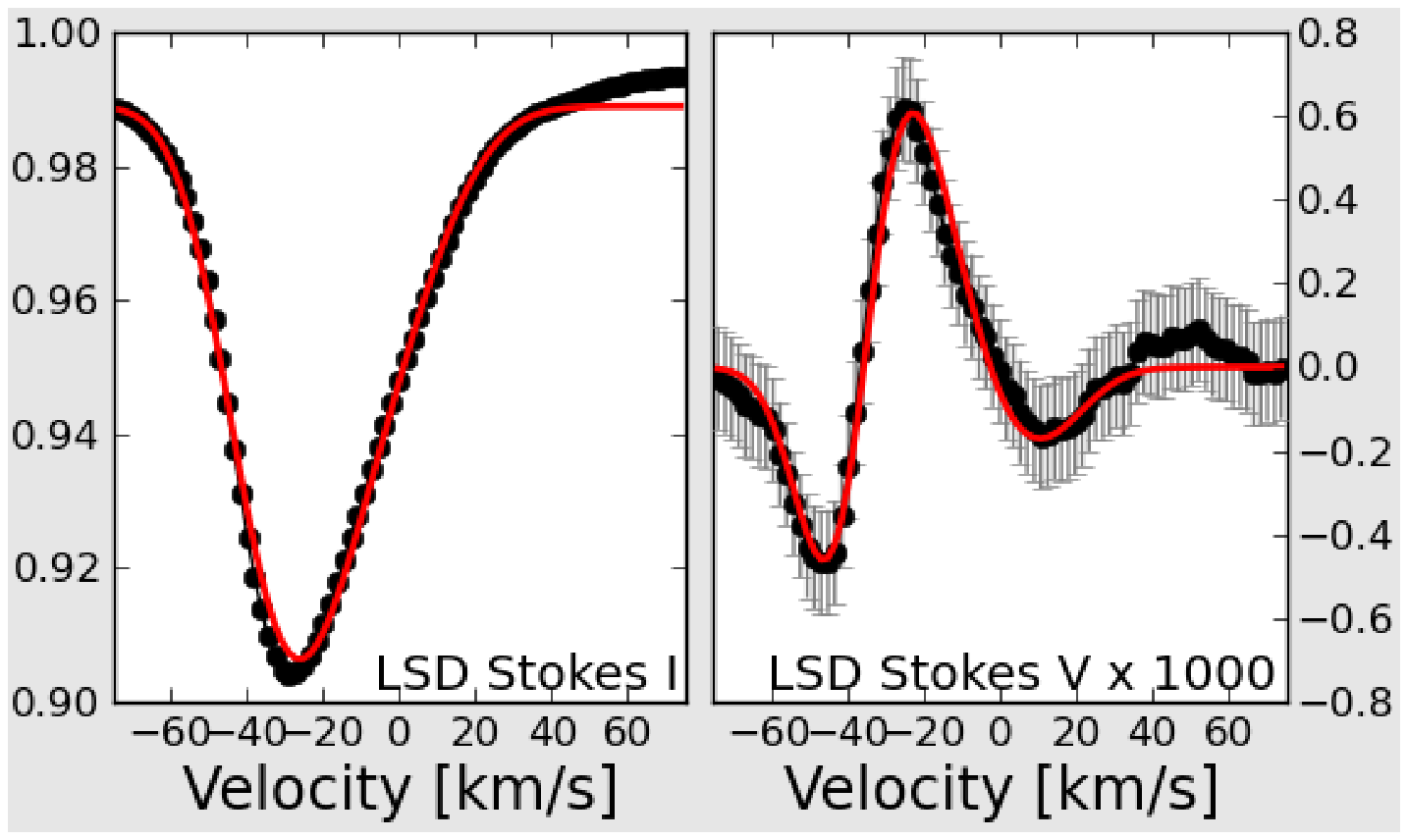}\\
\caption{Examples of the modelled surface (left), LSD Stokes I (middle) and V (right) profiles of $\beta$\,Cep fitted with pulsations and magnetic field, at two different phases.}
\label{Neiner_fig1}
\end{center}
\end{figure}

For the first time we thus modelled the intensity and Stokes V profiles of a
magnetic massive star, $\beta$\,Cep, taking into account its pulsations with the
new Phoebe 2.0 code (see Fig.~\ref{Neiner_fig1}), and we compared the results
with a standard oblique dipole model without pulsations. Without pulsations we
find i=89$^\circ$, $\beta$=51$^\circ$, B$_{\rm pol}$=389 G, while with
pulsations our preliminary results are i=70$^\circ$, $\beta$=50$^\circ$, B$_{\rm
pol}$=276 G. In particular, the resulting field strength seems significantly
lower when taking pulsations into account. 

\section{Taking magnetism into account in seismic studies}

The impact of a fossil magnetic field on rotation and mixing can be estimated following
two theoretical criteria: (1) the Spruit criterion \cite[(Spruit
1999)]{spruit1999}: Above a critical strength, the magnetic field freezes
differential rotation and mixing, and the field stays oblique. Otherwise the
structure adjusts to a symmetric configuration by rotational smoothing; (2) the
Zahn criterion \cite[(Zahn 2011, Mathis \& Zahn 2005)]{zahn2011,
mathiszahn2005}: The Lorentz force removes differential rotation along poloidal
field lines above a certain field strength and thus removes mixing.

For the magnetic pulsating B star V2052\,Oph, the Spruit and Zahn critical field
strengths are B$_{\rm crit}$=40 G and 70 G, respectively. In this star, the
measured polar field strength is B$_{\rm pol}$=400 G \cite[(Neiner et al.
2012)]{neiner2012}. Therefore we expect no mixing in V2052\,Oph. Indeed, a
seismic model of the pulsations of V2052\,Oph shows no overshoot in this star
\cite[(Briquet et al. 2012)]{briquet2012}. This example shows how a magnetic
field study can provide constraints for seismic modelling.

A magnetic field also produces splitting of the pulsation modes and a
modification of the amplitude of the pulsation modes. The split multiplet
depends both on the strength of the field and on its obliquity \cite[(see
Shibahashi \& Aerts 2000)]{shibahashi2000}. No magnetic splitting has been
identified so far in massive stars. However, the observation with CoRoT of
regular splittings in the hybrid B pulsator HD\,43317 \cite[(Papics et al.
2012)]{papics2012} and the recent discovery of a magnetic field in this star
\cite[(Briquet et al. 2013)]{briquet2013} make it an ideal candidate.

\section{Conclusions}

It is crucial to take pulsations into account when modelling the magnetic field
strength and configuration in pulsating massive stars. Moreover, knowing this
magnetic configuration provides important constraints on seismic modelling, in
particular it constraints the mixing, differential rotation and identification
of the modes.

\end{document}